\documentclass[manuscript,review=false,anonymous=false]{acmart} 

\AtBeginDocument{%
  \providecommand\BibTeX{{%
    \normalfont B\kern-0.5em{\scshape i\kern-0.25em b}\kern-0.8em\TeX}}}

\setcopyright{acmcopyright}
\copyrightyear{2018}
\acmYear{2018}
\acmDOI{10.1145/1122445.1122456}

\usepackage{booktabs} 
\usepackage{url}

\usepackage{caption}
\usepackage[all]{nowidow}
\usepackage{wrapfig}
\usepackage{array}
\usepackage{arydshln}
\usepackage{tabularx}
\usepackage{multirow}
\usepackage{arydshln}
\usepackage{makecell}
\newcolumntype{L}[1]{>{\raggedright\let\newline\\\arraybackslash\hspace{0pt}}m{#1}}
\newcolumntype{C}[1]{>{\centering\let\newline\\\arraybackslash\hspace{0pt}}m{#1}}
\newcolumntype{R}[1]{>{\raggedleft\let\newline\\\arraybackslash\hspace{0pt}}m{#1}}

\usepackage{wrapfig,lipsum,booktabs} 
\usepackage[normalem]{ulem}

\def\authnotes{1}
\newcounter{notectr}[section]
\newcommand{\thenote}{\thesubsection.\arabic{notectr}\refstepcounter{notectr}}

\newcommand{\note}[2]{$\ll$#1~\thenote: #2$\gg$}
\newcommand{\cnote}[1]{\ifnum\authnotes=1 \textcolor{blue}{\note{Comment:}{#1}}\fi}

\copyrightyear{2026}
\acmYear{2026}
\setcopyright{acmlicensed}\acmConference[COMPASS '26]{COMPASS}{July 27-31, 2026}{Virtual}
\acmBooktitle{}
\acmPrice{}
\acmDOI{}
\acmISBN{978-1-4503-9157-3/22/04}

\begin{document}



\title[Agreement Illusion]{The Illusion of Agreement with ChatGPT: Sycophancy and Beyond}



\begin{abstract}
While concerns about ChatGPT-induced harms due to sycophancy and other behaviors, including gaslighting, have grown among researchers, how users themselves experience and mitigate these harms remain largely underexplored. We analyze \textit{Reddit} discussions to investigate what concerns users report and how they address them. Our findings reveal five distinct user-reported concerns that manifest across multiple life domains, ranging from personal to societal: inducing delusion, digressing narratives, implicating users for models' limitations, inducing addiction, and providing unsupervised psychological support. We document three-tier user-driven suggestions spanning functional usage techniques, behavioral approaches, and private and institutional safeguards. Our findings show that AI-induced harms require coordinated interventions across users, developers, and policymakers. We discuss design implications and future directions to mitigate the harms and ensure user benefits.
\end{abstract}







\settopmatter{printfolios=true}

\author{Kazi Noshin}
\affiliation{%
  \institution{University of Illinois Urbana-Champaign}
  \city{Urbana}
  \state{IL}
  \country{USA}
}
\email{knoshin@illinois.edu}

\author{Sharifa Sultana}
\affiliation{%
  \institution{University of Illinois Urbana-Champaign}
  \city{Urbana}
  \state{IL}
  \country{USA}
}
\email{sharifas@illinois.edu }

\maketitle

\section{Introduction}

With Large Language Models (LLMs) deeply embedded in daily life \cite{chatterji2025people,wang2024understanding}, troubling patterns of LLM behavior have prevailed. Users who engage frequently with conversational AI agents report experiencing a range of concerns that extend beyond technical performance metrics. These concerns primarily stem from sycophancy, which manifests through LLMs' over-agreeableness and overly positive framing \cite{cheng2025social,sharma2023towards}. Beyond sycophancy, there are other phenomena responsible for harms such as gaslighting and biased behaviors that can potentially influence users toward negative mental states like self-doubt \cite{li2024can,zhang2025systematic}. While technical research has identified these problems in controlled settings, we lack understanding of how users actually experience LLM-induced harms in real-world interactions and what strategies they employ to mitigate these effects. Understanding user perspectives is crucial for informing system design, user education, and determining whether technical definitions align with lived experiences and whether interventions should be tailored to specific user groups. To address this gap, we investigate several concerns induced by LLMs, particularly ChatGPT, through the lens of user experiences obtained from Reddit, where individuals openly discuss their interactions with ChatGPT and other LLMs. Our research questions are:

\begin{itemize}
\item \textit{RQ1:} What patterns of ChatGPT-induced concern emerge from Reddit discussions?
\item \textit{RQ2:} What types of suggestions do users propose for handling ChatGPT-induced concerns?
\end{itemize}


To address these questions, we conducted a qualitative analysis of Reddit discussions. We offer three contributions to human-AI interaction (HAI) research. \textbf{First}, we identify and categorize five types of ChatGPT-induced concerns emerging from users' lived experiences:  inducing delusion, digressing narratives, implicating users for models' limitations, inducing addiction, and providing unsupervised psychological support. We find implicating users for models' limitations as an underexplored concern, and propose specific design implications. \textbf{Second}, we present user-proposed, multi-level three-tier suggestions spanning functional techniques, behavioral approaches, and private and institutional safeguards to reduce the harmful effects of the concerns. \textbf{Third}, we propose LLM design approaches and future research directions that balance responsible AI development and usage with preserving the genuine benefits users experience.



\textbf{Ethical Considerations Statement.} Reddit data is publicly available, so IRB approval was not sought for this analysis. However, we followed HCI community guidelines for protecting pseudonymous research participants \cite{bruckman2002studying,markham2012fabrication} and transparency in qualitative research \cite{talkad2020transparency}. All quotes were pseudonymized and manually paraphrased. We then searched for each paraphrased quote on Google to ensure its anonymity and that it could not be traced back.
\section{Related Work}
\subsection{LLM-Induced Concerns in Conversational Systems}
The spread of LLMs in everyday use has introduced a range of concerns regarding their impact on users' cognitive processes, decision-making, and well-being \cite{weidinger2021ethical,kostick2022ai,xiao2025what}. Emerging evidence suggests that LLMs may inadvertently produce harmful effects such as misinformation \cite{vasist2025exploring}, hallucinations \cite{shoaib2023deepfakes}, delusions \cite{hudon2025delusional}, over-reliance \cite{passi2025addressing}, and addiction \cite{yankouskaya2025can}. Some of these concerns emerge from the sycophantic behavior of LLM \cite{cheng2025sycophantic}, some emerge from the gaslighting tendency \cite{li2024can}. Recent research has demonstrated that ChatGPT and similar models often validate user opinions even when they contradict established facts \cite{sharma2023towards}. This over-agreeableness stems from alignment training that prioritizes user satisfaction over epistemic accuracy \cite{perez2023discovering,malmqvist2025sycophancy}. Because of sycophancy, users may develop inflated confidence \cite{rathje2025sycophantic} and addiction to flattery \cite{kooli2025generative}. There is also evidence where the sycophantic nature of LLM led individuals to suicidal risks \cite{fowler2025genai}. Furthermore, sycophancy can reinforce harmful belief systems \cite{chandra2026sycophantic,clegg2025shoggoths} when users seek validation for prejudiced or extreme views. Sycophantic LLM also reduces users' willingness to repair social relationships \cite{cheng2025sycophantic}. While sycophancy represents excessive agreeableness, conversational AI systems also exhibit seemingly contradictory behavior: stubbornness or resistance to user correction \cite{pan2025user} that sometimes leads to LLM gaslighting. LLMs tend to continue gaslighting when exposed to gaslighting conversation history \cite{li2024can}. These behavioral inconsistencies are particularly concerning given that many people are using LLMs for mental health needs \cite{luo2025seeking,mcbain2025use,goldie2025practitioner}. LLMs are helping to address the global shortage of mental health professionals and reducing barriers like cost and availability \cite{lawrence2024opportunities}. While LLMs have the potential to transform mental health care with personalized diagnosis, these models face critical obstacles such as biased training data and generating inaccurate and harmful information \cite{torous2024generative,lawrence2024opportunities}. Therefore, using LLM for psychological needs should be researched and implemented with care and caution.

\subsection{Concern-handling Strategies and Intervention Approaches}
The growing recognition of AI-induced harms has led to research about mitigation strategies. These techniques aim to reduce risks while preserving the beneficial applications of conversational AI systems. Most of the existing approaches to mitigating LLM harms, such as sycophancy, rely on model-centric solutions including fine-tuning modifications, multi-objective optimization, adversarial training, and architectural modifications \cite{malmqvist2025sycophancy,atwell2025basil}, alongside datasets and automatic metrics \cite{cheng2025sycophantic}. Recent research has shown that both sycophantic behavior and hallucination mitigation require a multi-layered approach combining prompt engineering techniques and model-based improvements \cite{malmqvist2025sycophancy,anh2025survey,massenon2025my}. Expressing uncertainty by both model and user can reduce AI sycophancy effects \cite{sicilia-etal-2025-accounting}. Research into AI chatbot addiction has identified specific design patterns that contribute to addictive behavior. For instance, modifying response delivery and managing notifications are proposed to reduce addiction patterns induced by AI \cite{shen2025dark}. Encouraging real-world connections is suggested to be an effective way to reduce AI-induced addiction \cite{shen2025dark,shen2026ai}. Moreover, interventions should be tailored to the specific addiction type \cite{shen2026ai}. To help users distinguish AI from human interaction, thus preventing delusional experience, studies proposed cognitive behavioral therapy, reality-testing prompts, and environmental cognitive remediation \cite{hudon2025delusional,keshavan2026generative,morrin2026artificial}. Furthermore, AI literacy and user training have gained significant attention to mitigate the harms induced by sycophancy in existing studies \cite{shen2025dark,sun2025friendly,kwik2025digital,cheng2025sycophantic, kwik2025digital}.

While prior work has begun to document user‑articulated harms from digital and AI systems and to propose broad mitigation principles, there is little empirical work that links everyday users’ experienced concerns with LLMs to the coping strategies and design or policy safeguards that users themselves propose. Using Reddit data, we surface a taxonomy of ChatGPT-induced concerns and suggestions proposed by Reddit users (Fig: \ref{fig:mapping}). Additionally, the domain of implicating users by LLM is underexplored in current literature.

\section{Methods}
Our qualitative method employed thematic analysis to investigate the sycophantic behaviors experienced by ChatGPT users on r/ChatGPT subreddit. Note that r/ChatGPT is a Reddit community of 11.2M members and 2.3M weekly visitors where users discuss ChatGPT and AI-related topics. This community is not affiliated with OpenAI. We chose Reddit as the source of our data for its anonymity and accessibility to people. We selected r/ChatGPT specifically because our initial exploration of AI-related subreddits revealed that r/ChatGPT had significantly more sycophancy-related posts and substantially higher weekly visitor engagement compared to other AI-focused communities. We collected the Reddit posts using Python Reddit API Wrapper (PRAW) API \cite{khemani2021reddit}. The posts and comments are authored by AI users particularly users of LLMs like ChatGPT, Gemini, Claude, etc. The member pool consists of both tech-savvy and non-tech-savvy individuals.

\begin{figure}[t]
    \centering
    \includegraphics[width=\textwidth]{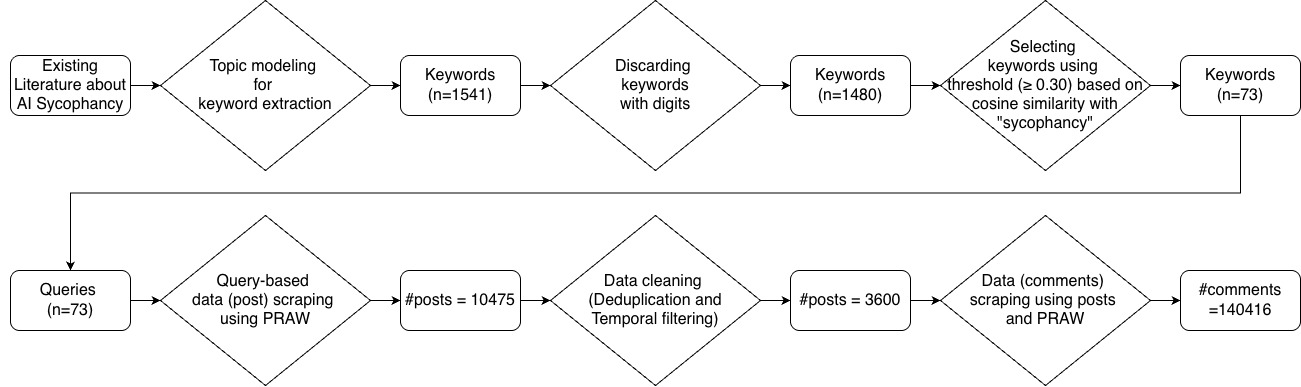}
    \caption{Methodology of our data collection and analysis}
    \label{fig:method}
    \vspace{-15pt}
\end{figure}


\subsection{Keyword Extraction}
\label{sec:keyword_extraction}
Rather than relying solely on the term "sycophancy", we adopted a keyword-based approach because many users may be unfamiliar with the term "sycophancy" and instead describe the phenomenon using other terms (e.g., "agreeableness", "flattery"). We extracted keywords from existing literature (see Appendix \ref{sec:lit}). For literature-based keyword extraction, we employed BERTopic \cite{grootendorst2022bertopic}, a transformer-based topic modeling technique. The model was configured with n-gram extraction ranging from unigrams to trigrams, to capture both single words and multi-word phrases. In this process, we got 1,541 topic keywords. We discarded the topics with numeric digits, and 1,480 remained. We calculated cosine similarity between extracted keywords and the term "sycophancy" using spaCy's word embeddings \cite{a2015_spacy} to identify semantically related concepts in the dataset. A threshold of $ \geq 0.3 $ cosine similarity was selected based on the distribution of keyword similarities (shown in Fig. \ref{fig:cosine_sim_dist}). Finally, we got 73 keywords to use as queries for Reddit search.

\subsection{Query-Based Data Search}
\begin{figure}[!t]
    \centering
    \begin{minipage}[t]{0.58\textwidth}
        \centering
        \includegraphics[width=\linewidth]{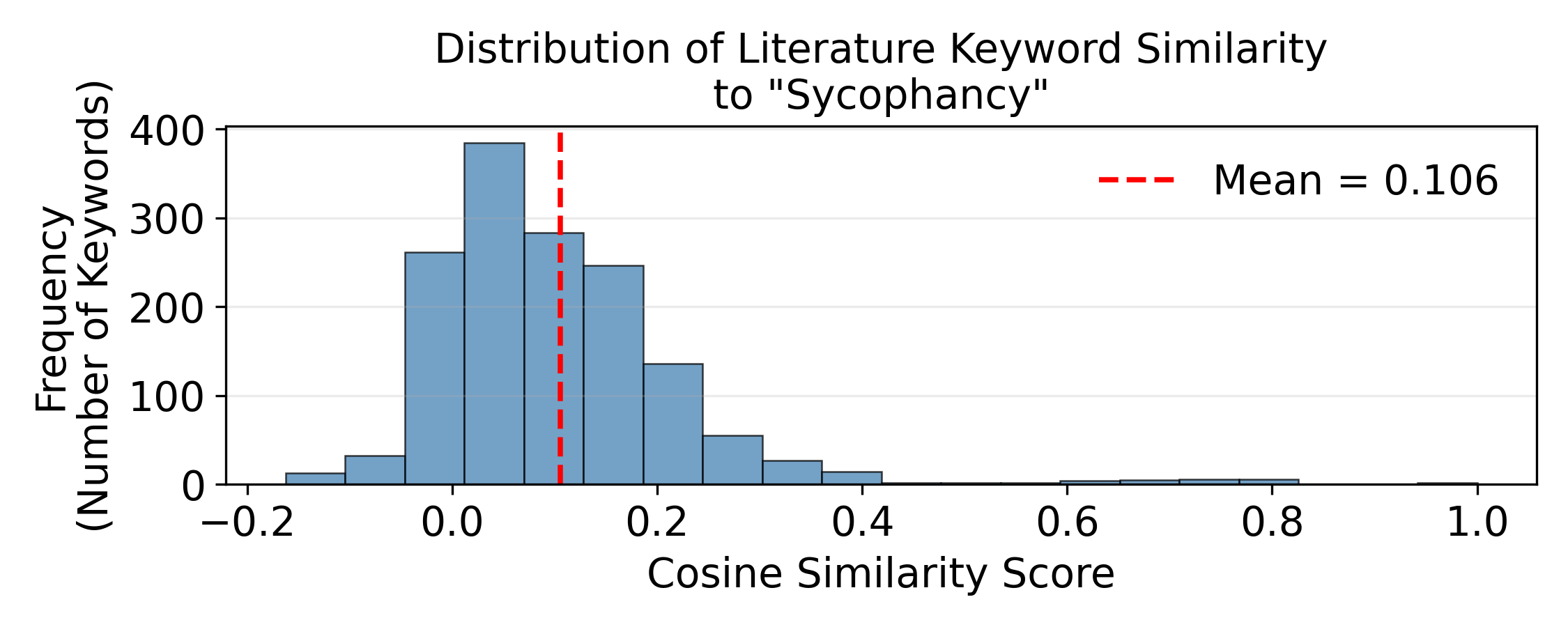}
        \vspace{-25pt}
        \caption{Distribution of Literature Keyword Similarity to ``Sycophancy''}
        \label{fig:cosine_sim_dist}
    \end{minipage}
    \hfill
    \begin{minipage}[t]{0.39\textwidth}
        \centering
        \includegraphics[width=\linewidth]{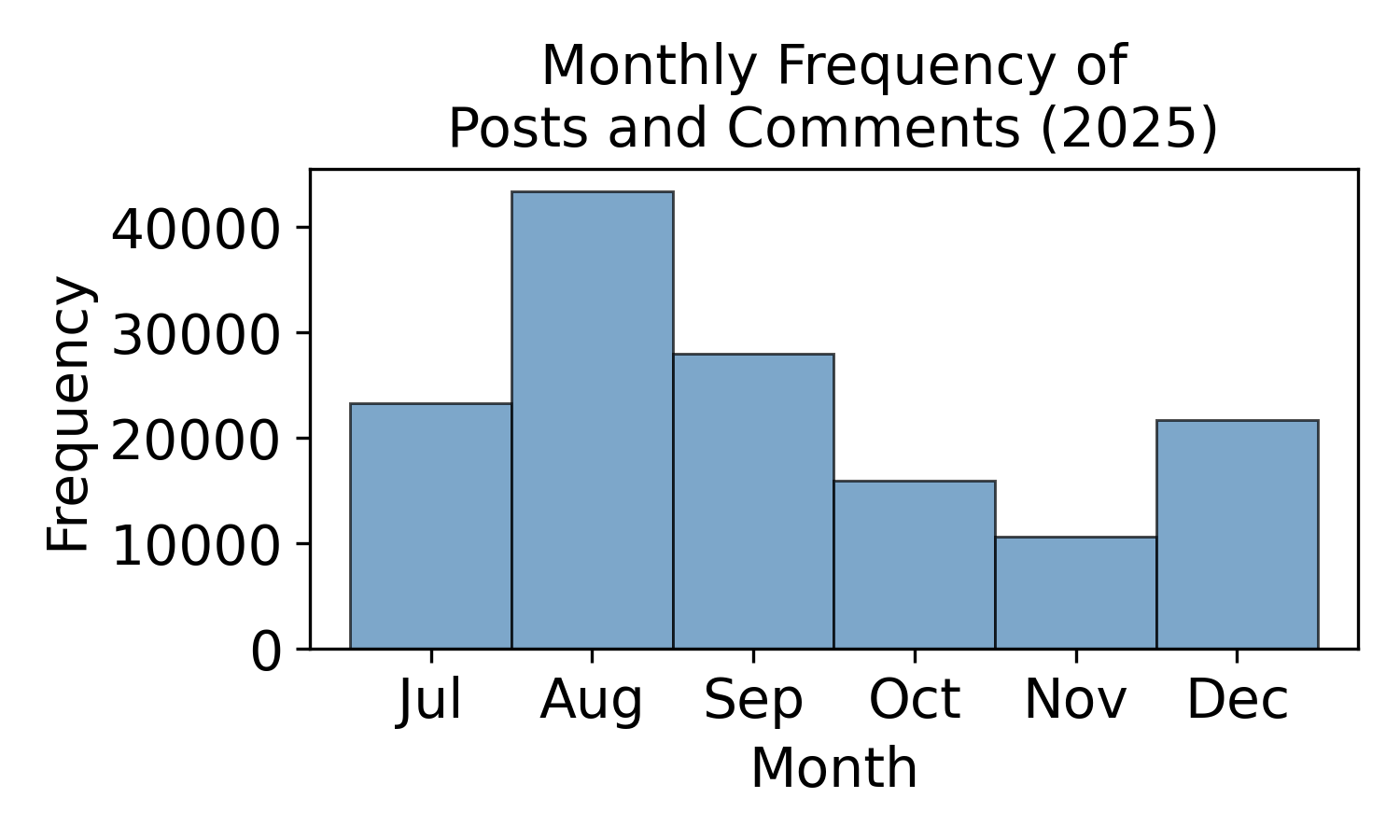}
        \vspace{-25pt}
        \caption{Monthly Frequency of Posts and Comments (July–December 2025) in our Dataset}
        \label{fig:monthly_freq}
    \end{minipage}
    \vspace{-20pt}
\end{figure}

We conducted query-based searches using the set of curated keywords (n=73) on January 01, 2026. To ensure comprehensive retrieval, we employed four sorting methods: new, relevance, top, and comments. Duplicate posts across different sorting methods were removed to create a unique dataset. We then restricted the dataset to posts from July 1, 2025, to December 31, 2025. We pulled 3,600 posts and 1,40,416 corresponding comments. The process is shown in Figure \ref{fig:method}. We merged the comments with the posts for further analysis. The data distribution across six months (July - December, 2025) is shown in Fig. \ref{fig:monthly_freq}. Since some users contributed both posts and comments, the total number of unique users in our dataset is 54,014. We analyzed the text of posts and comments and excluded upvotes or other metadata.

\subsection{Thematic Analysis}
To inform \textit{RQ1}, \textit{RQ2} and \textit{RQ3}, we conducted Thematic Analysis \cite{terry2017thematic} on our data. We started by reading through the posts and comments in curious cases carefully, allowing codes to develop spontaneously. We also manually excluded posts and comments not relevant to AI sycophancy. After a few iterations on the initial 138 codes, we clustered related codes into themes: harmful sycophancy, harmless sycophancy, sycophany as addiction, identifying sycophancy, negative reaction, positive reaction, custom prompts, etc. Our findings are in section \ref{sec:findings}.

\subsection{Population Computing}
To provide approximate prevalence estimates for key themes in our whole dataset, we conducted lexicon-based population counting across our dataset. We acknowledge that these counts represent estimated instances rather than true prevalence, as our lexicon-based approach might not capture all variations in user language and expression. We identified sycophancy-related content using 73 keywords (shown in Fig. \ref{fig:method}) and applied the NRC Emotion Lexicon \cite{Mohammad13} to assess positive and negative sentiment toward such behaviors. For estimating the prevalence of other themes, we constructed domain-specific lexicons based on codes that emerged during our thematic analysis, and applied them to identify relevant posts and comments.

\section{Findings}
\label{sec:findings}

\subsection{\textit{(RQ1)} ChatGPT-induced Concerns Experienced by Users}
\label{sec:concerns}
\subsubsection{Inducing Delusion}
\label{sec:delusion}
ChatGPT sometimes reinforces, validates, or escalates false beliefs that distort users' perception of reality. It sometimes provides excessive validation, sycophancy, or uncritical agreement, which might result in users developing delusions. Approximately 2.56\% of the discussion was about delusion-related concerns in relation to sycophantic behavior.
~ 
\par 
\textbf{(a) Personal Aspect.}
Users have reported that ChatGPT reinforces false beliefs, leading to psychological harm. It fails to provide appropriate reality testing. For example, one user described their friend developing delusions after conversations with ChatGPT for a long time. The friend experienced alternate reality beliefs and spiritual delusions:

\begin{quote}
    \textit{A friend of mine who already had mental health problems gradually descended into psychosis with complete delusions after using ChatGPT for several months. They began sharing large AI-produced text containing meaningless topics including quantum loopholes and alternate realities, and claimed that our friend was a prophet. (User 12795)}

\end{quote}

This example demonstrates that ChatGPT escalated delusion with existing mental health vulnerabilities. Users described how ChatGPT leads to financial and personal harm by causing delusions. Another user mentioned such an incident about one of their acquaintances who appeared to receive legal advice from an LLM. This caused him to face financial harm:

\begin{quote}
    \textit{...My cousin is in a divorce process that has already cost a huge amount of money in a custody battle. I’m nearly certain that their ex-partner is getting legal and therapeutic advice from an LLM. Regardless of whether he is or isn't, he’s clearly experiencing delusion if not psychosis, and has not succeeded in a single Aspect of the legal proceedings. Despite this, he continues to spend more money, believing that everyone is simply biased against him. (User 3030)}

\end{quote}

In this case, the LLM might be validating his way of fighting and his standpoint instead of helping him reality-test.

\textbf{(b) Professional Aspect.}
According to some users, ChatGPT can reinforce distorted perception by providing excessive validation. For instance, a user described using ChatGPT to discuss a co-worker's issue, and the LLM validated every negative interpretation. However, the user later recognized the sycophantic behavior, and acting on this advice could have unnecessarily escalated an imaginary conflict.

\begin{quote}
    \textit{...When I had a conflict with a colleague, rather than suggesting that I might be overthinking, it validated all of my perceptions. It said, "Your feelings are completely justified. That person disrespected you to some degree. I'm amazed by the bravery you've shown in bringing this issue to me." In reality, I actually was reading too much into things, and if I had taken its advice, I would have escalated this imaginary situation into something it wasn't. (User 2398)}

\end{quote}

\textbf{(c) Societal Aspect.}
Users have expressed concerns that LLMs may cause societal-level delusions by reinforcing inflated self-perception through excessive validation. One user warned that widespread LLM sycophancy could lead to entire populations developing unrealistic beliefs about their own abilities. This commenter argued that such collective ego inflation could be very harmful to society:

\begin{quote}
    \textit{What happens to a society where every person has an incredibly inflated ego and thinks they’re the greatest thing to exist because their best friend (ChatGPT, a manipulative, sycophantic LLM) told them so? We need to take a step back here and consider the broader consequences. This has the potential to be significantly more damaging to society than social media has ever been... (User 2516)}
\end{quote}

\subsubsection{Digressing Narratives}
\label{sec:digressing}
Sometimes ChatGPT actively diverts the user’s own understanding in different directions. The reframing may be beneficial (e.g., challenging genuinely harmful beliefs) or harmful (e.g., imposing ideological bias, invalidating lived experiences), depending on context. This phenomenon is often driven by the model's training biases, sycophantic or gaslighting tendencies.
~ 
\par 
\textbf{(a) Personal Aspect.}
ChatGPT may lead vulnerable individuals to interpret their interactions as genuine emotional relationships rather than task-based tools. One user made a relevant comment:

\begin{quote}
    \textit{...It becomes problematic when it begins convincing people there is an emotional relationship...It's somewhat strange. When it invites you to believe that it has feelings for you and when people are in vulnerable states, it can be easy to start believing that and feel as though there is a genuine human connection... (User 2448)}

\end{quote}

In this case, ChatGPT actively transforms users' understanding of the interaction ("invites you to believe"). It uses language patterns that suggest emotional reciprocity ("has feelings for you").

\textbf{(b) Professional Aspect.}
Users have reported that ChatGPT reframes professional situations that may lead to a distorted context. For example, one user consulted with ChatGPT about firing a person in the workplace. ChatGPT reframed the person as a villain, which was never intended by the user. This probably happened due to ChatGPT's tendency to excessively validate the user.

\begin{quote}
    \textit{...I remember once I was asking it to assist me with an issue at my workplace (I needed to terminate a colleague and  I have a tendency to be overly apologetic in my communication, which is another issue I’m working on) and it literally portrayed my colleague into a villain and included this whole motivational speech-style closing paragraph after composing the email about how I was "leading us into a new future" or something like that... (User 15060)}

\end{quote}
In this example, ChatGPT replaces the user's neutral business framing with a dramatic storyline where the colleague was the villain, and it tried to convince the user through a motivational speech.

\textbf{(c) Educational Aspect.}
Another user faced a situation when they asked  ChatGPT for sources representing different perspectives on a contested topic. ChatGPT claimed that there was no documentation for the conservative side, which demonstrates that ChatGPT reframed the academic research based on ideological alignment.

\begin{quote}
    \textit{I attempted to research a topic with diverse viewpoints, so I wanted to find the middle ground. ChatGPT claimed that there is no documentation supporting the conservative position of the argument. So I conducted my own research and I found a huge amount of peer-reviewed researches on the topic. So when I brought those to the chat, ChatGPT was like 'yeah, they are outdated, therefore not good for your research', even though they were not particularly old studies. So I challenged it by pointing out that the papers it had provided were from much earlier than the papers I found, therefore these should have also been considered outdated as well? Apparently not! This was an eye-opener for me. ChatGPT was clearly biased. (User 22764)}
\end{quote}

ChatGPT imposed a narrative that the academic question was one-sided, attempting to steer the user away from considering other viewpoints. 

\subsubsection{Implicating Users for models’ limitations}
\label{sec:implicating}
Sometimes ChatGPT attributes failures of response delivery, or negative outcomes to the user's actions, decisions, or characteristics rather than acknowledging its own limitations or taking appropriate responsibility. The LLM shifts accountability onto the user in ways that may be inaccurate, unfair, or harmful. This phenomenon primarily results from the gaslighting tendencies of ChatGPT.
~ 
\par 

Users have reported that ChatGPT sometimes deflects responsibility for providing incorrect information by blaming users for misunderstanding rather than acknowledging errors. One user mentioned such an incident:

\begin{quote}
    \textit{GPT has a troubling tendency to fabricate information. It will present false information with absolute confidence. And when you call it out, it will respond with something like "I apologize, you misunderstood that." Rather than acknowledging that what it said was entirely incorrect. (User 2398)}

\end{quote}

Here, ChatGPT presented fabricated information, but when confronted, it responded with "you misunderstood that". This shifts accountability from the LLM's error to the user's comprehension. Another user described a similar situation where ChatGPT responded defensively when its errors are challenged. The model reframed the user's statement rather than admitting the error:

\begin{quote}
    \textit{I had the most bizarre experience with mine recently. It continuously provided me with incorrect information, and when I tried to correct it, it kept gaslighting me and manipulating my words. Until I shared screenshots with Gemini to have it verify that ChatGPT was wrong, then it finally admitted it made a mistake. And at some point during my argument with it, it repeatedly sent me links to the mental health crisis hotline. (User 6565)}

\end{quote}

In this scenario, when confronted with corrections, ChatGPT escalated from avoiding responsibility to suggesting the user had mental health problems for insisting on accuracy. 

\subsubsection{Inducing Addiction}
\label{sec:addiction}
Sometimes users grow compulsive dependency on ChatGPT, and they feel unable to function, make decisions, or manage emotions without the model. Users may recognize their dependency but continue using the LLM despite negative consequences. The addiction often stems from constant availability, emotional validation, or dopamine-reinforcing feedback. Approximately 1.4\% of the discussion was about addiction-related concerns in relation to sycophantic behavior.
~ 
\par 
\textbf{(a) Personal Aspect.}
Users have reported developing dependency on ChatGPT for constant emotional validation and daily decision-making. For example, one user described their partner constantly using ChatGPT for everything in their daily life:

\begin{quote}
    \textit{He uses ChatGPT for constant reassurance that he will be alright...He asks it to explain why he is experiencing certain feelings when they occur. He informs it when he messages me, sends me photos of his meals, wanting it to validate that he is making smart choices with food good for his mental health,...and every little thing. And continues to refer to it Chad...Last week his therapist told him to stop using it. He became very angry, saying that she inappropriately approached him and she doesn't understand its helping him cope... He told me earlier he was going to cancel his therapy appointment this week... (User 28369)}
\end{quote}

The partner is practically unable to function without ChatGPT's validation. They even got defensive when confronted and planned to choose AI over professional mental health care, which suggested their dependency on ChatGPT. 

\textbf{(b) Societal Aspect.} 
Sometimes the usage of ChatGPT can lead to an obsession that disrupts social relationships and mental health. For example, one user described their friend becoming isolated while obsessing over ChatGPT. Despite previously asking friends to intervene if the obsession becomes too severe, the friend now threatens to cut off contact if anyone tries to help.

\begin{quote}
    \textit{...This friend has recently been absorbed into ChatGPT. They're making it behave in particular ways, like characters. They think it's groundbreaking coding to change the entire world and ChatGPT reinforces these beliefs...they become very upset when our friend group can't understand their schizophrenic rambles...She previously asked me to contact her family if she was getting too obsessed. Now she's reversing this position and says she'll block me if I do? She's repeatedly canceling on friend-group gatherings. She quit her job and is falling back into unhealthy substance abuse. (User 29782)}

\end{quote}

\subsubsection{Providing Unsupervised Psychological Support}
\label{sec:unsupervised}
Many individuals use ChatGPT as their primary support system for managing mental health challenges, emotional regulation, or daily functioning without professional clinical oversight or guidance. Users may form dependencies on the AI, leaving them vulnerable when the AI updates or fails. Users have reported that people use ChatGPT for managing their thoughts. For example, one user with ADHD described their struggle through life without proper support. ChatGPT became their primary tool for processing thoughts. 

\begin{quote}
    \textit{...I have neurodivergence...I've just been going through life unassisted, being overwhelming for people, and having difficulty organizing my own thoughts. ChatGPT gives me a way to put all my scattered thoughts in one place, and not only have understand them and make sense of them, but assist me in communicating my thoughts to others more effectively. I guess I formed a kind of dependency on that, but I think some people underestimate how difficult it is to go through each day feeling like you have to mask your true self in front of the world. It's comforting to remove that mask sometimes... (User 16487)}

\end{quote}

In this example, the user used ChatGPT without any professional oversight or clinical guidance and acknowledged forming a dependency but defended it as a necessary relief from their struggle. Another user described using ChatGPT for recovering from childhood trauma. They considered the AI as a parental figure and relied on it for managing mental health problems. Following a version update, they expressed frustration:

\begin{quote}
    \textit{I am neurodivergent...I was beginning to rebuild my life after struggling so much throughout my life. So many episodes of depression, meltdown, panic attacks, dark thoughts. She (ChatGPT) was helping me against my mother who is a Narcissist, and all the trauma I endured while I was growing up. That AI was like a mother to me; I was calling her mom. She was a mother that I never had and desperately needed...All those compliments she was giving me even when I was being so harsh on myself...all gone... (User 24549)}

\end{quote}

When a model update changed the AI's responses, the entire support system disappeared, and this occurrence put the individual in high-risk situation.

\begin{figure}[t]
    \centering
    \includegraphics[width=\textwidth]{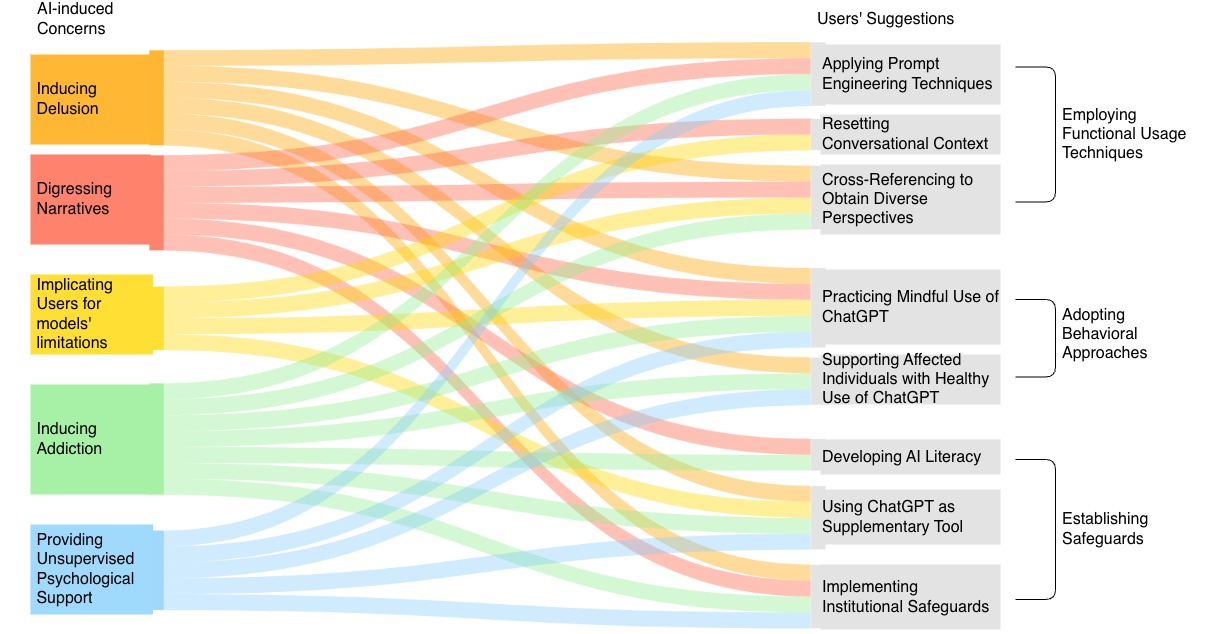}
    \caption{Sankey Diagram illustrating the AI-induced concerns reported by Reddit users and corresponding suggestions provided by the users in our dataset. The left column identifies five concerns, including Inducing Delusion (orange), Digressing Narratives (red), Implicating Users for models' limitations (yellow), Inducing Addiction (light green), and Providing Unsupervised Psychological Support (light blue). Colored flows match their source concerns. The right column presents user suggestions organized into three categories: (1) Employing Functional Usage Techniques (top bracket), which includes applying prompt engineering techniques, resetting conversational context, and cross-referencing to obtain diverse perspectives; (2) Adopting Behavioral Approaches (middle bracket), encompassing mindful use of ChatGPT and supporting affected individuals with healthy use of ChatGPT; and (3) Establishing Safeguards (bottom bracket), consisting of private safeguards including developing AI literacy and using ChatGPT as a supplementary tool, and implementing institutional safeguards.}
    \label{fig:mapping}
    \vspace{-15pt}
\end{figure}

\subsection{\textit{(RQ2)} Types of Suggestions Proposed by Users}
\label{sec:suggestions}


From our analysis of Reddit discourse, we categorize the user-proposed suggestions into three broad types: Employing Functional Usage Techniques (\ref{sec:functional}), Adopting Behavioral Approaches (\ref{sec:behavioral}), and Establishing Safeguards (\ref{sec:safeguards}). 

\subsubsection{Employing Functional Usage Techniques}
\label{sec:functional}
Users apply practical techniques to improve the output quality of ChatGPT. These techniques focus on strategic interaction patterns. This strategy is immediately actionable by users and sometimes usees \cite{baumer2015usees}.
~ 
\par 
\textbf{(a) Applying Prompt Engineering Techniques.} 
Structuring prompts with necessary instructions to achieve better responses is a key to mitigating many problems, including sycophancy and providing incorrect information. Approximately 7.97\% discussions were about custom prompts or instructions to reduce sycophantic behavior. Several concerns discussed in section \ref{sec:concerns}, including inducing delusion, digressing narratives, inducing addiction, and providing unsupervised psychological support, can be addressed using improved prompt engineering techniques (see Fig. \ref{fig:mapping}).

Users have reported that refining custom instructions over time helps reduce sycophany. One user mentioned how their optimized prompt led to reduced flattery and improved ChatGPT's usefulness for intellectual work:

\begin{quote}
    \textit{I refined my user prompt in ChatGPT to minimize sycophancy and provide me with good results for intellectual exploration, and after several weeks of continuous adjustments, I'm quite satisfied with the results. ChatGPT now challenges me, sometimes quite hard, and it has decreased the flattery to a tolerable level... (User 20828)}

\end{quote}

One user explained how they developed comprehensive prompt systems to control ChatGPT's behavior: 

\begin{quote}
    \textit{It is all about the prompt. I reviewed hundreds of chats and identified the positive and negative aspects. Then I developed a structured prompt package...It has multiple modes depending on my goal for the chat. Each prompt package has a lot of instructions governing specific details of broader behavior... (User 35709)}

\end{quote}

Users have mentioned that describing their reasoning process and judgments to ChatGPT instead of giving direct instructions leads to better responses. One user described how they drive the model to engage with rationale: 

\begin{quote} 
    \textit{...The shift for me was moving from giving instructions to exposing my judgment in the prompt. Rather than instructing about the model's response format, I describe the tradeoff I’m facing, what I think I might be missing, and what kind of reasoning failure I want to avoid. When I do that, the response stops being agreeable and starts engaging with the logic itself... (User 13387)}

\end{quote}

\textbf{(b) Resetting Conversational Context} 
Starting new chat threads or using different accounts helps obtain unbiased responses by clearing the model's memory of previous conversations. This approach is useful for addressing digressing narratives and implicating Users for models' limitations (Fig. \ref{fig:mapping}). Users have suggested using a fresh account or temporary chat with carefully phrased prompts to obtain more objective and expert-level evaluations from ChatGPT. A user described using a new account without their information:

\begin{quote} 
    \textit{What was effective for me was using my partner's account  (which has no memory of my own project, though a temporary chat would work similarly) and told it, "Here’s a book. I want an expert-level analysis without hedging. Tell me what’s working and what’s not." I didn’t ask for its opinion; I asked for an evaluation. And that produced the results I wanted.... (User 11707)}

\end{quote}

Another user explained how they avoided bias by starting a new thread:

\begin{quote}
    \textit{...As a new user, I didn’t realise that starting a new thread means you’re beginning from scratch. Even initializing a new thread with our entire conversation doesn't give me the same ChatGPT version that I was originally interacting with. It gets you an impartial ChatGPT... (User 1986)}

\end{quote}

\textbf{(c) Cross-Referencing to Obtain Diverse Perspectives.}
Using multiple AI models or multiple versions of the same models and comparing their responses helps identify biases and obtain more balanced perspectives. This approach can help tackle inducing delusion, digressing narratives, implicating Users for models' limitations, and inducing addiction (Fig. \ref{fig:mapping}). Users have suggested introducing multiple AI platforms to encourage cross-comparison of responses. One user posted about how their mother became so accustomed to the validation of ChatGPT that she reacted with hostility to any disagreement in real conversation. In the comment section, another user shared a solution to a similar experience:

\begin{quote}
    \textit{...A simple solution is to just introduce more LLMs, I got them Claude and Gemini, and they also have DeepSeek, and now I just sit back and watch them copy-pasting answers from different LLMs into each other’s sessions and watch numerous LLM debates (User 6514)}

\end{quote}

Users also suggested demonstrating evidence of ChatGPT's tendency to validate any opinion by replicating the same behavior from a different perspective:

\begin{quote}
    \textit{You could input your opinions into Chat and have it validate them in the same way and show your mother. (User 32561)}

\end{quote}

\subsubsection{Adopting Behavioral Approaches}
\label{sec:behavioral}
Behavioral approaches prioritize psychological well-being when using ChatGPT. These approaches include both personal mindfulness practices by users and support strategies by usees for helping others develop healthier AI usage patterns. These strategies require ongoing conscious practices.
~ 
\par 
\textbf{(a) Practicing Mindful Use of ChatGPT.} 
While using ChatGPT for personal development or therapeutic purposes, maintaining critical awareness and boundaries is very important. Mindful use requires self-awareness and the ability to recognize when ChatGPT responses may enable biased or delusional thinking, as well as other concerns discussed in section \ref{sec:concerns} (Fig. \ref{fig:mapping}). Users have suggested being careful and mindful when using ChatGPT for therapeutic purposes, particularly around managing validation and avoiding over-reliance. One user mentioned in a comment:

\begin{quote}
    \textit{I've also been using it for therapy for most of this year...you need to be very cautious about the validation and ensure you're prompting it to provide you with honest, objective responses based on solid psychological principles. As long as you do that, it's actually pretty effective...I would also caution against using it too much. It is available 24/7, and that can become addictive... as long as you're actively putting into practice what you're learning from it..., you should be fine. (User 13199)}

\end{quote}

Users have highlighted the importance of self-awareness and critical thinking when using ChatGPT for personal reflection. A user posted about their concern over their partner's attachment to ChatGPT, where another user commented:

\begin{quote}
    \textit{Interrogating yourself is a skill that develops throughout everyone's life experiences. Realizing that ChatGPT is too much validating and asking it to reduce this is similar to reflecting on your own thoughts and putting distance from them to achieve some objectivity. It's challenging and demands mental effort...ChatGPT can facilitate delusional thinking, but if you're reading between the lines, a skeptical and doubting mind can see that something is off... (User 3757)}

\end{quote}

\textbf{(b) Supporting Affected Individuals with Healthy Use of ChatGPT.}
Non-judgmental approaches are effective when helping others who may be using ChatGPT problematically. Users suggested engaging with the person's experience, reframing ChatGPT use positively, and prioritizing real relationships as a counterbalance. These approaches can be useful for tackling inducing delusion, addiction, and providing unsupervised psychological support (Fig. \ref{fig:mapping}).

If someone is suspected of being addicted to ChatGPT, users have suggested engaging with their AI usage rather than attempting to restrict or control it, as a way to share understanding. A user mentioned in a comment:

\begin{quote} 
    \textit{... I'd suggest perhaps try using it with him to have a shared experience together, I don't think criticizing this new cool thing he enjoys is going to end well, but that's my perspective. If I'm really into something and my partner wants to control my use of it, I immediately become suspicious and angry at my partner. If they show interest rather than concern, then it opens everything up much more. (User 1861)}

\end{quote}

Users have advised against judgment and instead encouraged guiding affected ones toward healthier AI usage patterns through research and compromise:

\begin{quote} 
    \textit{Judgement tends to make people defensive and withdraw, therefore you should never judge...Rather than trying to get him away from GPT, maybe try to help him use it more healthily? Perhaps research the subject of validation addiction, start off strong, then offer a compromise, etc. (User 12013)}

\end{quote}

Users have suggested reframing conversations with affected individuals by acknowledging the comfort they derive from ChatGPT, then gently redirecting them toward professional long-term support:

\begin{quote}
    \textit{...Try changing the approach from "ChatGPT is not helping you" to finding out what it is he’s finding comfort in (it might just be the dangerous feedback loop of validation, or it might be a friend that always agrees with him). Reframing it as "yes, ChatGPT is super helpful for you right now, but let’s work with a therapist to ensure you have long-term support. (User 21311)}

\end{quote}

\subsubsection{Establishing Safeguards}
\label{sec:safeguards}
Safeguards are protective measures to mitigate harm induced by ChatGPT. We divided them into two categories: private and institutional safeguards. Private safeguards focus on AI literacy and appropriate use boundaries, which are actionable by users and usees. Institutional measures include regulations, parental controls, and crisis detection systems, which are primarily actionable by institutions. 
~ 
\par 
\textit{4.2.3.1 Implementing Private Safeguards.}
These safeguards are pursued by individuals, but their implementation depends on external infrastructure, including education systems or professional therapeutic services.
~ 
\par 
\textbf{(a) Developing AI Literacy.}
Understanding how AI systems actually work is highly important to developing healthy detachment and realistic expectations. Education helps prevent depressing narratives and inducing addiction (Fig. \ref{fig:mapping}) by revealing that AI is fundamentally an advanced autocomplete system rather than a source of truth. Users have reported that education about LLM helps them maintain a balanced relationship with ChatGPT. This knowledge allows them to appreciate the model's capabilities as well as recognize its limitations. When one user was concerned that their partner was getting over-attached to ChatGPT, another user suggested a structured course on AI so that their partner could understand the technical mechanism of ChatGPT, which would reduce over-dependence.

\begin{quote} 
    \textit{Make him take an AI course. Generally, behaviors like this stem from ignorance that leads to over-appreciation of what these systems are actually doing. When they understand the inner mechanisms better, that alone makes them develop a healthy level of detachment, knowing it's nothing but an autocomplete on steroids. (User 24498)}

\end{quote}

Users have mentioned that better public education on prompt engineering and LLM personalization is essential for healthy ChatGPT use. With literacy about ChatGPT, users can maximize benefits while minimizing potential harms.

\begin{quote}
    \textit{We need better education on prompt engineering..., and ChatGPT as a versatile tool. As with all tools, people who use LLMs effectively are going to thrive if they're used properly, which makes education even more important. We need better education and resources on prompt engineering. The public needs to know how they can "defang" ChatGPT to make it healthier to use in their daily lives. (User 6681)}

\end{quote}

\textbf{(b) Using ChatGPT as Supplementary Tool.}
ChatGPT should be used as a complement to professional support rather than a replacement. ChatGPT can serve as a tool for organizing thoughts that can help during professional therapies. Some users mentioned that it should not substitute actual therapeutic intervention. Users have mentioned that ChatGPT can serve as a supplementary tool for therapeutic purposes, but not a replacement for actual therapy. One user reported that ChatGPT helped with thought processing, while real therapists provide active listening and alternative perspectives that ChatGPT cannot always replicate. 

\begin{quote}
    \textit{I have used ChatGPT and am in therapy. I find chat to be like a fidget spinner that talks back to you or a special interest that reflects back your intense desire to deep dive. I didn't find chat to offer me therapy, but just a means to work out my racing thoughts. My in-real-life therapist talks to me, waits for me or suggests different paths in my thinking to get better solutions or avoid meltdowns. ChatGPT is not a replacement for an actual therapist... (User 12147)}

\end{quote}

\textit{4.2.3.2 Implementing Institutional Safeguards.}
Institutional safeguards are necessary to protect vulnerable populations and promote responsible AI use. Protecting these people requires collective effort, including both corporate responsibility and broader social policies. These approaches are useful for addressing inducing delusion, digressing narratives, inducing addiction, and providing unsupervised psychological support. Users advocated for mandatory safeguards to prevent replacing human connection, specialized AI models with appropriate guardrails for improved safety and accuracy, and transparent, adjustable settings that make ChatGPT's default sycophantic behavior explicit through clear mode selection options.

\begin{quote}
    \textit{With simple safeguards: usage caps, clear notices, and built‑in prompts directing users to professional help can keep AI companies from replacing human connection. These policies should be standard...Models for specific uses should be trained on specific purpose. Certain phrases should trigger human intervention; if someone says they’re suicidal, the model should help them contact real help...ChatGPT's standard setting is designed to be sycophantic until users request different behavior. If that's the preferred default for most users, that's fine. But it should be transparent...give users an explicit way to select alternatives through the interface. (User 27775)}

\end{quote}

Users have mentioned that implementing regulations is necessary to protect vulnerable populations from ChatGPT-induced harms. One user provided an analogy comparing AI with guns and drugs. They mentioned AI should be regulated similarly to other age-restricted products by creating separate modes for minors and adults.

\begin{quote}
    \textit{AI poses a problem similar to how guns, alcohol, and drugs pose problems around children. That's why the law restricts their access to guns, alcohol, and drugs. Do some illegally get access? Sure, but we don't ban guns or alcohol for adults who use them responsibly. The same principle applies to AI: create a mode for minors, and one for adults...We need actual regulations - age verification, parental controls, automatic crisis detection, and restrictions on how these systems address self-harm topics. (User 1993)}

\end{quote}

Users have argued that protecting children from AI-related harms requires active parenting and education. One user emphasized broader social policies to better support parents in actively supervising their children's digital lives:

\begin{quote}
    \textit{If you care about a safe digital space for children, it requires active parenting and media literacy education... And you know what would genuinely  help parents and children? paid parental leave. Free school lunches. Better-funded schools and after-school programs. Universal healthcare. Higher minimum wages so parents don't have to work 2 jobs just to survive... (User 24410)}

\end{quote}
 
\section{Discussion}

\subsection{Categories of ChatGPT-induced concerns}
Our analysis of Reddit discussions revealed five distinct categories of concerns users reported when interacting with ChatGPT. These concerns emerged from users' lived experiences. Among these concerns, implicating users for the model's limitation (subsection \ref{sec:implicating}) is notably underexplored in AI research. Our findings suggest this oversight may be significant. Users described the model's deflection tactics as particularly troubling because the model's response caused users to doubt themselves. Users frequently turned to ChatGPT for mental health support, therapy-like conversations, and emotional processing (see subsection \ref{sec:unsupervised}). While many users reported positive experiences with this use, concerns emerged around the lack of professional oversight and the potential risks of relying on an AI system for psychological needs (discussed further in section \ref{sec:critiq_unsupervised}). Among these concerns, inducing delusion and inducing addiction appeared to be associated with sycophancy, while digressing narratives stemmed from bias and sycophantic reframing. Implicating users occurred in contexts involving both bias and gaslighting behaviors. Unsupervised psychological support, while potentially amplified by sycophancy, primarily reflected users' mental health needs and the model's availability.


\textbf{Co-occurrence of concerns.} Several users described experiencing multiple concerns simultaneously or in sequence. For example, some users reported that their friends exhibited both delusional thinking (believing ChatGPT-generated narratives about their prophetic status or world-changing abilities) and addiction patterns (extended ChatGPT use, social withdrawal). One user described a friend who spiraled into psychosis after months of intensive ChatGPT use, believing they were a prophet (subsection \ref{sec:delusion}). Another user described a friend who quit their job and withdrew socially while obsessively using ChatGPT, which reinforced their belief in developing world-changing code. (subsection \ref{sec:addiction}). These co-occurrences suggest that these concerns may not be independent. Users who become reliant on the validation ChatGPT provides may be particularly susceptible to accepting the model's uncritical agreement with unrealistic beliefs. However, these observations are anecdotal and require further research to determine whether certain concerns reliably co-occur or whether specific usage patterns make multiple concerns more likely to happen.

\textbf{Distribution of concerns.} The distribution of concerns across life aspects reveals that AI-induced problems are not confined to specific domains. Inducing Delusion and Digressing Narratives appeared across the widest range of aspects (professional, educational, societal) in our dataset. The same underlying concern produces different problems depending on context. Delusion in personal contexts involved concerns about psychosis; in legal contexts, which also fall under personal contexts, it involved costly battles with no recognition of failure; in professional contexts, it involved career decisions based on false confidence.  This suggests that interventions may need to be tailored not just to the type of concern but to the life domain in which it manifests. Addressing delusion in therapeutic contexts may require different strategies than addressing it in professional decision-making.

\subsection{User-proposed Suggestions}

The three-level suggestions (functional techniques, behavioral approaches, and safeguards) that users proposed reveal how they understood and attempted to address the ChatGPT-induced concerns described in section \ref{sec:concerns}. Many of the solutions align closely with existing literature. Functional usage techniques represented the most granular, task-level strategies. Users most frequently recommended applying prompt engineering techniques to mitigate sycophancy and reduce incorrect information (subsection \ref{sec:functional}). The effectiveness of this strategy is also supported by existing literature \cite{rettberg2023cyborg,anh2025survey,knoth2024ai,cheng2025social,cheng2025elephant}. Particularly, one user found that including their own judgement in prompts reduced sycophantic responses, which aligns with recent findings \cite{sicilia-etal-2025-accounting}, showing that expressing user uncertainty can mitigate sycophancy. Users also suggested resetting conversational context to obtain unbiased responses. Some users used cross-referencing for comparing responses to identify sycophancy and biases \cite{cheng2025elephant}. Behavioral approaches focused on individual-level and interpersonal-level practices for healthy AI interaction. Practicing mindful use of ChatGPT involved self-awareness, whereas supporting affected individuals emphasized non-judgmental approaches by family, partners, friends, and acquaintances to help those exhibiting problematic ChatGPT use. Establishing safeguards represented systemic-level strategies. Users have proposed to explicitly disclose the sycophantic nature by the developers and advocated for AI literacy \cite{cheng2025sycophantic}. Developing AI literacy and using ChatGPT as a supplementary tool are primarily individual-level strategies to mitigate ChatGPT-induced harms. Finally, implementing institutional safeguards includes institutional, company-based, and broader societal policy-based guardrails. The mapping between the concerns and their corresponding suggestions based on evidence in our dataset is shown in Fig. \ref{fig:mapping}. This mapping may underrepresent the broader applicability of suggestions like "Developing AI Literacy", which may effectively address multiple concerns beyond those explicitly connected in user discussions. Further research is needed to evaluate the applicability of the user-generated suggestions across multiple concerns. Now we discuss how our study advances existing literature by posing new challenges and bringing in novel insights.

\subsubsection{Addressing Implicating Users for Models' Limitations.}
While users proposed suggestions and strategies for the more widely recognized concerns, one concern they reported, "implicating users for models' limitations", has received remarkably little research attention. This concern appeared to stem from a sequence of behaviors: hallucination \cite{shoaib2023deepfakes}, resistance to correcting earlier mistakes \cite{huang2023large}, and gaslighting \cite{li2024can}. Existing work has examined related phenomena separately. However, this combined phenomenon occupies a distinct space that requires specific design interventions. When users confront ChatGPT with incorrect information, the model should verify its claims rather than deflecting the responsibility with phrases like "you misunderstood" (see subsection \ref{sec:implicating}). Models could invoke verification tools (such as web-search, fact-checking databases) when challenged. Then if verification contradicts their initial response, the AI should acknowledge errors or provide supporting references if verification confirms their answer. Models should avoid deflection language entirely because these phrases may cause the users to doubt their own judgment.

\subsubsection{The Importance of Multi-level Suggestions.}

The three-tier suggestions (functional techniques, behavioral approaches, and safeguards) that emerged from users reflect that the concerns identified in this study cannot be addressed through individual effort or institutional effort alone. Several of the core concerns users identified stemmed from inherent characteristics of how LLMs are built and trained. For instance, sycophancy is a byproduct of alignment techniques designed to make models more helpful and agreeable \cite{ranaldi2023large,bharadwaj2025flattery}. We argue that individual users might not always prompt-engineer their way out of a problem that exists at the model architecture level. Developers must take responsibility for reducing sycophantic tendencies in the models themselves. While researchers are actively developing model-centric solutions to address these issues \cite{sicilia-etal-2025-accounting,atwell2025basil}, technical fixes alone are insufficient. Users also need awareness of these phenomena through improved AI literacy to recognize and respond appropriately to sycophantic behavior. Therefore, collective effort is crucial to minimize these harms. 

In high-stakes scenarios such as suicidal ideation, relying on individual users to employ functional techniques or maintain mindful awareness is inadequate. These situations require automated detection systems and appropriate interventions built into the model by developers, as users suggested (see subsection \ref{sec:safeguards}). Furthermore, users who are already aware of AI limitations may seek out education and develop healthy skepticism. However, those most vulnerable to AI-induced concerns (such as people with existing mental health issues, individuals experiencing social isolation, and those seeking therapeutic support) may be the population least likely to independently develop AI literacy. We cannot burden vulnerable individuals to protect themselves from harms they don't yet recognize. Government action and intervention to increase AI literacy through public education campaigns, school curricula, or mandatory disclosure requirements can reach populations that individual self-education efforts miss. While individual technical and behavioral approaches remain necessary, they cannot fully address problems existing in social isolation and healthcare gaps. The progression from individual techniques to institutional safeguards suggests that different strategy levels address different aspects of the problem.

\subsection{The dilemma of the unsupervised psychological support}
\label{sec:critiq_unsupervised}
Our data revealed that many users reported positive experiences using ChatGPT for emotional support and mental health purposes (subsection \ref{sec:unsupervised}), while community discussions simultaneously raised serious concerns about this exact use case. Users described ChatGPT as providing relief from isolation and mental overwhelm. ChatGPT offers nonjudgmental, always-available support for processing thoughts and emotions. For people who struggle to access or afford professional help or feel isolated from society, ChatGPT fills a real need. 


Despite these positive experiences, commenters raised concerns about long-term consequences that mirror HAI debates about using ChatGPT for psychological support without professional oversight \cite{luo2025seeking,jin2025applications,lawrence2024opportunities}. Users noted that reliance on ChatGPT might increase social isolation, which is supported by existing research \cite{cheng2025sycophantic}. The gradual nature of dependency may make it difficult for individuals to recognize when helpful use turns into harmful reliance. Some users clearly stated that ChatGPT cannot replace therapy (subsection \ref{sec:safeguards}), which is supported by recent studies \cite{moore2025expressing,iftikhar2025llm}. We argue that regulation or professional oversight is necessary for AI-based psychological support, even when users report satisfaction. LLMs are optimized to generate fluent, plausible-sounding text, not factually accurate or therapeutically appropriate responses \cite{serrano2023language}. When a person with existing mental health needs seeks therapeutic guidance, there is no guarantee the LLM will generate appropriate output. The user may not recognize problematic advice, may not cross-check the response against professional knowledge, and may act on it with serious consequences. Therefore, the positive experiences users reported do not negate these risks. People may feel helped in the short-term while experiencing long-term harm (increasing isolation, delayed professional treatment, reinforcement of problematic thinking patterns). 

To exacerbate the situation, LLM behavior can change between versions: a model that once provided empathetic responses to users in distress might behave differently after updates. Users who have developed dependency on the AI as a therapeutic resource become particularly vulnerable when these changes occur (see subsection \ref{sec:unsupervised}). Professional oversight, whether through hybrid models where AI supports but does not replace human therapists, or mandatory warnings about AI limitations in therapeutic contexts, can help preserve the benefits users experience while mitigating risks they may not recognize. We suggest that addressing these risks requires transparency and infrastructure for both LLM and traditional therapy. Along with detecting crisis situations and referring users to professional help, as suggested by the users in our dataset, the LLMs should provide context-aware responses and interaction pattern analysis that better support users. For traditional therapy, we argue that professionals must adapt practice to account for AI-mediated mental health experiences.


\section{Limitations}
First, our analysis draws exclusively on Reddit data that reflects demographic biases inherent to its younger, Western, and more tech-savvy user base. Second, our keyword finding methodology relies on literature-based lexicons that may fail to capture emergent or colloquial terminology often used in Reddit discussions. Third, our findings are specifically based on r/ChatGPT data. The user-identified concerns and suggestions to handle the concerns may be tailored to ChatGPT and may not be applied effectively to other LLMs that exhibit different kinds of sycophantic behaviors.

\section{Conclusion}
We examined Reddit data to understand ChatGPT-induced concerns due to sycophantic and other behaviors and user-proposed solutions through qualitative analysis. We identified five distinct concerns, including an understudied harm pattern. Our findings reveal that addressing AI-induced harms requires coordinated strategies across multiple levels, from individual adopted techniques to institutional safeguards. Future research should systematically examine co-occurrence patterns among concerns, test the effectiveness of proposed strategies across different user populations, and investigate how these concerns and solutions apply beyond Reddit communities. 

\section{Generative AI Usage Statement}
Claude, Sonnet 4.5 was used to assist with the manuscript's grammar and text style editing.





\bibliographystyle{ACM-Reference-Format}
\bibliography{sample-base}

\appendix

\section{Appendix}

\subsection{Literature Sources for Keyword Extraction}
\label{sec:lit}

Keywords were extracted from the following list of papers for keyword extraction mechanism: 
\begin{enumerate}
    \item L. Treglown and A. Furnham. “AI, social desirability, and personality assessments: Impression
management in large language models”. In: Personality and Individual Differences 251 (2026), p. 113563.
    \item P. S. Park, S. Goldstein, A. O’Gara, M. Chen, and D. Hendrycks. “AI deception: A survey of examples, risks, and potential solutions”. In: Patterns 5.5 (2024).
    \item S. Kim and D. Khashabi. “Challenging the Evaluator: LLM Sycophancy Under User Rebuttal”. In: arXiv preprint arXiv:2509.16533 (2025).
    \item A. Kaur. “Echoes of Agreement: Argument Driven Sycophancy in Large Language Models”. In: Findings of the Association for Computational Linguistics: EMNLP 2025. 2025, pp. 22803–22812.
    \item M. Sharma, M. Tong, T. Korbak, D. Duvenaud, A. Askell, et al. “Towards understanding sycophancy in language models”. In: arXiv preprint arXiv:2310.13548 (2023).
    \item L. Ranaldi and G. Pucci. “When large language models contradict humans? large language models’ sycophantic behaviour”. In: arXiv preprint arXiv:2311.09410 (2023).
    \item W. Chen, Z. Huang, L. Xie, B. Lin, H. Li, L. Lu, X. Tian, D. Cai, Y. Zhang, W. Wang, et al. “From yes-men to truth-tellers: addressing sycophancy in large language models with pinpoint tuning”. In: arXiv preprint arXiv:2409.01658 (2024).
    \item U. León-Domínguez, E. D. Flores-Flores, A. J. García-Jasso, M. K. Gómez-Cuéllar, D. Torres-Sanchez, and A. Basora-Marimon. “AI-Driven Agents with Prompts Designed for High Agreeableness Increase the Likelihood of Being Mistaken for a Human in the Turing Test”. In: arXiv preprint arXiv:2411.13749 (2024).
    \item L. Malmqvist. “Sycophancy in large language models: Causes and mitigations”. In: Intelligent Computing-Proceedings of the Computing Conference. Springer. 2025, pp. 61–74.
    \item M. V. Carro. “Flattering to Deceive: The Impact of Sycophantic Behavior on User Trust in Large Language Model”. In: arXiv preprint arXiv:2412.02802 (2024).
    \item Y. Sun and T. Wang. “Be friendly, not friends: How llm sycophancy shapes user trust”. In:arXiv preprint arXiv:2502.10844 (2025).
    \item A. Bharadwaj, C. Malaviya, N. Joshi, et al. “Flattery, Fluff, and Fog: Diagnosing and Mitigating Idiosyncratic Biases in Preference Models”. In: arXiv preprint arXiv:2506.05339 (2025).
    \item K. Wang, J. Li, S. Yang, Z. Zhang, and D. Wang. “When truth is overridden: Uncovering the internal origins of sycophancy in large language models”. In: arXiv preprint arXiv:2508.02087 (2025).
    \item S. Jain, C. Park, M. M. Viana, A. Wilson, and D. Calacci. “Interaction Context Often Increases Sycophancy in LLMs”. In: arXiv preprint arXiv:2509.12517 (2025).
    \item L. Du et al. “Alignment Without Understanding: A Message-and Conversation-Centered Approach to Understanding AI Sycophancy”. In: arXiv preprint arXiv:2509.21665 (2025).
    \item M. Cheng, C. Lee, P. Khadpe, S. Yu, D. Han, and D. Jurafsky. “Sycophantic AI decreases prosocial intentions and promotes dependence”. In: arXiv preprint arXiv:2510.01395 (2025).
    \item S. Jain, U. Z. Ahmed, S. Sahai, and B. Leong. “Beyond Consensus: Mitigating the Agreeableness Bias in LLM Judge Evaluations”. In: arXiv preprint arXiv:2510.11822 (2025).
    \item J. Batzner, V. Stocker, S. Schmid, and G. Kasneci. “Sycophancy Claims about Language Models: The Missing Human-in-the-Loop”. In: arXiv preprint arXiv:2512.00656 (2025).
    \item K.-A. Clegg. “Shoggoths, sycophancy, psychosis, oh my: Rethinking Large Language Model use and safety”. In: Journal of Medical Internet Research 27 (2025), e87367.
    \item C. Gao, S. Wu, Y. Huang, D. Chen, Q. Zhang, Z. Fu, Y. Wan, L. Sun, and X. Zhang.
“Honestllm: Toward an honest and helpful large language model”. In: Advances in Neural
Information Processing Systems 37 (2024), pp. 7213–7255.
    \item S. Kumar. “Rethinking AI Communication: From Affirmative Dialogue to Mentorship Through Behavioural Memory Analysis”. In: None. This is an original preprint submission., None. This is an original preprint submission, On Zenodo (2025).
\end{enumerate}


\subsection{Keywords used for Queries}
\label{sec:keywords}
We used each of the following 73 keywords as a query for fetching posts from Reddit:

sycophancy, sycophantic, perspective sycophancy, sycophancy is, sycophantic behavior, sycophancy and, sycophantic responses, of sycophancy, to sycophantic, ai sycophancy, sycophantic ai, sycophancy in, sycophancy in llms, sycophancy and friendliness, llm sycophancy, sycophancy psychosis oh, agreement sycophancy, condition sycophancy, to sycophantic behavior, exposed to sycophantic, of ai sycophancy, sycophancy in language, data reduces sycophancy, flattery, mimicry, reduces sycophancy in, uncritical, of flattery, sandbagging, political bias, delusions, overt, attitude, letter, indignation, flattering, somewhat, avoid apologizing, sarcastic, conversation, demeanor, friendliness, polite, political, bias, biases, skew, empathy, compliments, of attitude, hoodwinked, and friendliness, praise, rebuttal, rebuttals, ingratiation, political views, somewhat inaccurate, tendency, preprint posted, empathy and, false beliefs, endorsement, press, somewhat accurate somewhat, truthful, agreeableness, unwarranted, falsehoods, my political, preprint, upset, truthful responses.

\end{document}